\documentclass[aps,pra,preprint,groupedaddress]{revtex4}

\usepackage{amssymb}
\usepackage{epstopdf}
\usepackage{graphicx}
\usepackage{bm}
\newcommand{\bra}[1]{\langle #1|}
\newcommand{\ket}[1]{|#1\rangle}

\begin{document}

\preprint{LA-UR-05-5519}

\title{Quantum Bayesian methods and subsequent measurements}


\author{Filippo Neri}
\thanks{Research sponsored by the US Department of Energy under contract W-7405-Eng-36}
\affiliation{Los Alamos National Laboratory, Los Alamos, NM 87545, USA.}


\date{\today}

\begin{abstract}
After a derivation of the quantum Bayes theorem, and a discussion of the
reconstruction of the unknown state of identical spin systems by 
repeated measurements, the main part of this paper treats the problem 
of determining the unknown phase
difference of two coherent sources by photon measurements.
While the approach of this paper is based on computing
correlations of actual measurements (photon detections),
it is possible to derive indirectly a probability distribution
for the phase difference. In this approach,
the quantum phase is not an observable, but a parameter
of an unknown quantum state. Photon measurements  determine a
probability distribution for the phase difference. The approach used
in this paper takes into account both photon statistics and the finite
efficiency of the detectors.

\end{abstract}


\maketitle

\section{Introduction}
The use of Bayes theorem in quantum mechanics is discussed. It is shown that the
quantum Bayes theorem follows from the ordinary quantum measurement theory, 
when applied to density operators that represent our a-priori knowledge of a system. 
The examples studied involve measurements on multiple copies of the same
(unknown) state.  The theorem is used to determine the unknown state by
successive measurements on several of the copies of the state.

The theorem is applied to quantum optics:
an idealized information-theoretic
description of propagating CW laser beams is treated in detail.
It is shown how photon detections
on part of the beams can be used to  determine the phase of the rest of the beams.
Explicit expressions are derived for the conditional probabilities of
detecting photons at different positions, given the numbers of photons
detected at different positions in the past.
The quantitative predictions could be used, in principle, to test
models of the quantum state of propagating laser beams.

I hope that the information-theoretic approach used in
this paper will be a useful contribution to the understanding of the
problem of quantum phase.

\section{Bayes Theorem}
Bayes theorem follows from the definition of conditional probability:
\begin{equation}
 P(A \vert B) = { {P (A \cap B)}\over{P(B)}}. \label{Bayes:0}
\end{equation}
Bayes theorem allows us to reverse probability relationships.
For instance, if the probability of the observation $B$, given the assumption
(theory?) $A$ is $P(B \vert A)$, then
the probability of $A$, after having observed $B$ is:
\begin{equation}
 P(A \vert B) = { {\Pi (A) P(B \vert A)}\over{\sum_{A^{\prime}} \Pi (A^{\prime})
 P(B \vert A^{\prime})}}. \label{Bayes:1}
\end{equation}
$\Pi(A)$ is the a-priori probability of $A$. The form of Eq.~(\ref{Bayes:1}) is the
most useful for analyzing classical data. The Bayes Theorem requires some
way to provide the a-priori probabilities (priors). In some cases the priors
can be supplied by symmetry arguments, in other cases, repeated applications
of Eq.~(\ref{Bayes:1}) will provide a result only weakly dependent
on the original priors. 

In quantum mechanics one has to go back to the form of Eq.~(\ref{Bayes:0}):
\begin{equation}
P(A \vert B) = {{{\rm{Tr}}(\hat{A}_{A} \hat{A}_{B} \hat{\Pi} \hat{A}_{B}^{\dagger}
\hat{A}_{A}^{\dagger})}
\over {{ \rm{Tr}}(\hat{A}_{B} \hat{\Pi} \hat{A}_{B}^{\dagger})}}, \label{QBayes}
\end{equation}
where $\hat{\Pi}$ is a density operator containing our a-priori knowledge of the
system being measured.  $\hat{A}_{B}$ and $\hat{A}_{A}$ are measurement
operators corresponding to $B$ and $A$.

If $\hat{A}_{B}$ and $\hat{A}_{A}$ commute with each other (for instance if they act on different
subspaces---as in all the following examples), then Eq.~(\ref{QBayes}) can be put in the form
\begin{equation}
P(A \vert B)= {{{\rm{Tr}}(\hat{\Pi} \hat{E}_{B} \hat{E}_{A})}\over {{ \rm{Tr}}(\hat{\Pi} \hat{E}_{B})}}.
\label{QBPOVM}
\end{equation}
Here $\hat{E}_{A}$ and $\hat{E}_{B}$ are the POVMs corresponding to the operators $\hat{A}_{B}$
and $\hat{A}_{A}$:
\begin{equation}
\hat{E}_{B} = \hat{A}_{B}^{\dagger} \hat{A}_{B}, \: \hat{E}_{A} = \hat{A}_{A}^{\dagger} \hat{A}_{A}.
\label{POVM}
\end{equation}
Under the same assumptions that led to Eq.~(\ref{QBPOVM}), one can derive the conditional
probability of measuring 
$A_1, A_2 \cdots A_n$, after having measured $B_1, B_2 \cdots B_m$:
\begin{equation}
P(A_1, A_2 \cdots A_n \vert B_1, B_2 \cdots B_m) =
 {{{\rm{Tr}}(\hat{\Pi} \hat{E}_{B_1} \hat{E}_{B_2} \cdots \hat{E}_{B_m} \hat{E}_{A_1}
 \hat{E}_{A_2} \cdots \hat{E}_{A_n})}\over {{ \rm{Tr}}(\hat{\Pi} \hat{E}_{B_1} \hat{E}_{B_2} 
 \cdots \hat{E}_{B_m} )}}. \label{QBPOVMN}
\end{equation}
Note that the quantum Bayes ``theorem'' as expressed here in Eq.~(\ref{QBPOVM}) and
(\ref{QBPOVMN}),  
is not a new principle, but a consequence of the ordinary quantum measurement theory. 
The only special assumptions that will be made, will be the choices of the a-priori density
operators $\hat{\Pi}$.

\section{Spin Systems}
An example of the use of Eq.~(\ref{QBayes}) is a system of $N+M$ qubits (see Refs.~\cite{Sch:01} 
and \cite{Buz:98}). 
Following Ref.~\cite{Sch:01} we assume the a-priori density operator
\begin{equation}
\hat{\Pi}_{\rm{Qbits}} = {\int \! {d^3}\vec{x}  \Pi(\vec{x})\left({{\hat{1}+\vec{x}\cdot{\hat{\vec{\sigma}}}}
\over2}\right)^{\otimes{(N+M)}}}, 
\label{qubits}
\end{equation}
where the vector $\vec{x}$ has components $x$, $y$ and $z$. The vector operator
$\hat{\vec{\sigma}}$
has components $\hat{\sigma}_x$, $\hat{\sigma}_y$ and $\hat{\sigma}_z$ 
(the Pauli operators).
Eq.~(\ref{qubits}) means that all $M+N$ qubits are in the same state, represented
by the same density
operator, which is unknown, except for the a-priori probability distribution $\Pi(\vec{x})$, 
which is normalized so that its integral is equal to one.
The form of Eq.~(\ref{qubits}) follows from the quantum de Finetti representation theorem.
(For information about the quantum de Finetti theorem see
Refs.~\cite{Hud76} and \cite{Cav02}.)

The measurement operator corresponding to measuring $M_x$ times $\hat{\sigma}_x$
(on different qubits), $M_y$ times $\hat{\sigma}_y$ and $M_z$ times $\hat{\sigma}_z$
($M_x+M_y+M_z=M$), with the result $\sigma_x=+1$ obtained $M_x^{+}$ times and
$\sigma_x=-1$ obtained $M_x^{-}$ times (with $M_x^{+}+M_x^{-}=M_x$), etc. is
\begin{equation}
\hat{A}_{M_x^+, M_x^-,M_y^+, M_y^-,M_z^+, M_z^-}  = 
{\hat{1}}^{\otimes N}\otimes {\left({{{\hat{1}+\hat{\sigma}_x}}\over2}\right)}^{{\otimes}M_x^{+}}
\cdots \otimes {\left({{{\hat{1}-\hat{\sigma}_z}}\over2}\right)}^{{\otimes}M_z^{-}} +
\rm{Permutations}. \label{PMM}
\end{equation}
Similarly, the measurement operator corresponding to measuring $N_x$ times $\hat{\sigma}_x$,
$N_y$ times $\hat{\sigma}_y$ and $N_z$ times $\hat{\sigma}_z$ ($N_x+N_y+N_z=N$),
with the result $\sigma_x=+1$ obtained $N_x^{+}$ times and $\sigma_x=-1$
obtained $N_x^{-}$ times (with $N_x^{+}+N_x^{-}=N_x$), etc. (also all on different qubits) is
\begin{equation}
\hat{A}_{N_x^+, N_x^-,N_y^+, N_y^-,N_z^+, N_z^-}  = 
{\left({{{\hat{1}+\hat{\sigma}_x}}\over2}\right)}^{{\otimes}N_x^{+}}
\cdots \otimes {\left({{{\hat{1}-\hat{\sigma}_z}}\over2}\right)}^{{\otimes}N_z^{-}}
\otimes {\hat{1}}^{\otimes M}  + \rm{Permutations}. \label{PNN}
\end{equation}
Inserting Eqs.~(\ref{qubits}), (\ref{PMM}) and (\ref{PNN}) into Eq.~(\ref{QBayes}),
we get a result for the probability of measuring $\sigma_x=+1$ $N_x^{+}$ times,
$\sigma_x=-1$ $N_x^{-}$ times, etc., after having measured $\sigma_x=+1$ $M_x^{+}$
times, $\sigma_x=-1$ $M_x^{-}$ times, etc.
(with each measurement performed on a different qubit):
\begin{eqnarray}
\lefteqn{P(N_x^+, N_x^-,N_y^+, N_y^-,N_z^+, N_z^- \vert
M_x^+, M_x^-,M_y^+, M_y^-,M_z^+, M_z^- ) = }
\nonumber  \\
& &  {{(N_x^+ \! +\! N_x^-)!} \over {N_x^+ ! N_x^- !}} {{(N_y^+\! +\! N_y^-)!} \over
 {N_y^+ ! N_y^- !}} {{(N_z^+\! + \! N_z^-)!} \over {N_z^+ ! N_z^- !}}  \times \nonumber \\ 
 & &  {{\int \! {d^3}\vec{x} \Pi(\vec{x}) ({1+x\over2})^{M_x^+ \! + \! N_x^+} 
 ({1-x\over2})^{M_x^- \! + \! N_x^-} ({1+y\over2})^{M_y^+ \! + \! N_y^+} 
 ({1-y\over2})^{M_y^- \! + \! N_y^-} ({1+z\over2})^{M_z^+ \! + \! N_z^+} 
 ({1-z\over2})^{M_z^- \! + \! N_z^-} } \over  {\int \! {d^3}\vec{x} \Pi(\vec{x}) 
 ({1+x\over2})^{M_x^+} ({1-x\over2})^{M_x^-} ({1+y\over2})^{M_y^+} 
 ({1-y\over2})^{M_y^-} ({1+z\over2})^{M_z^+} ({1-z\over2})^{M_z^-} }    }. 
 \label{spin_big} 
\end{eqnarray}
Eq.~(\ref{spin_big}) is an obvious consequence of the discussion of Ref.~\cite{Sch:01}.
I wrote it down explicitly to illustrate the symmetry between $M$s and $N$s in the numerator.
This symmetry has consequences for the naive use of the relation
\begin{equation}
\lim_{M_x \rightarrow \infty} {\left({1+x\over2}\right)^{{M_x}^+}}
 {\left({1-x\over2}\right)^{{M_x}^-}} = 
 \delta(x-{{{M_x}^+ - {M_x}^-}\over{{M_x}^+ + {M_x}^-}}), \label{del?}
\end{equation}
up to a normalization factor. Eq.~(\ref{del?}) is only applicable in Eq.~(\ref{spin_big})
if the $M$s are large compared to the $N$s. So a large, but finite number of measurements
will determine the state accurately enough only if the number of successive measurements
is small compared to the number of measurements used to determine the state.

\subsection{Exact Solutions}
To illustrate this point, we will give a complete treatment of the case in which the spin
measurements are all along the same axis. We will give an analytical solution with the
prior of Ref.~\cite{Buz:98}. In this case, the integral of Eq.~(\ref{spin_big}) is constrained
to the surface of the unit sphere, with uniform weight. (This is---of course---not the only
possible pure-state prior: one could use a non-uniform weighting of the unit sphere surface.) 
The a-priori probability of measuring $M_x^+$ times $\sigma_x = +1$
and $M_x^-$ times $\sigma_x = -1$ is 
\begin{equation}
P(M_x^+, M_x^-) =  {{(M_x^+ \! +\! M_x^-)!} \over {M_x^+ ! M_x^- !}}
{\int_0^{\pi} d{\theta} \int_0^{2\pi} {{\sin{\theta} d{\phi}}\over{4\pi}} 
\left({1+\cos{\theta}\over2}\right)^{M_x^+} \left({1-\cos{\theta}\over2}\right)^{M_x^-}}. 
\label{PMMS}
\end{equation}
Substituting $x$ for $\cos{\theta}$, we can put Eq.~(\ref{PMMS}) in the form
\begin{equation}
P(M_x^+, M_x^-) = 
 {{(M_x^+ \! +\! M_x^-)!} \over {M_x^+ ! M_x^- !}}  {\int_{-1}^{1} 
 {{d{x}}\over{2}} \left({1+x\over2}\right)^{M_x^+} \left({1-x\over2}\right)^{M_x^-}}. \label{PMMC}
\end{equation}
The conditional probability of getting $N_x^+$ times $\sigma_x = +1$
and $N_x^-$ times $\sigma_x = -1$---after having measured $M_x^+$
times $\sigma_x = +1$ and $M_x^-$ times $\sigma_x = -1$---is
\begin{equation}
P(N_x^+, N_x^- \vert M_x^+, M_x^-) =   
{{(N_x^+ \! +\! N_x^-)!} \over {N_x^+ ! N_x^- !}}
{{\int_{-1}^{1}  {{d{x}}\over{2}} ({1+x\over2})^{M_x^+ \!+\! N_x^+} 
({1-x\over2})^{M_x^- \!+\! N_x^-}} \over {\int_{-1}^{1}  {{d{x}}\over{2}} 
({1+x\over2})^{M_x^+} ({1-x\over2})^{M_x^-}}}.  \label{PNNMMC}
\end{equation}
Using the generating function method, one can show that $P(M_x^+ ,M_x^- )$ has the value
\begin{equation}
P(M_x^+ ,M_x^- ) = {1\over{M_x^+ \!+\! M_x^- \!+\! 1}}. \label{INT+-}
\end{equation}
This means that all possible outcomes of the $M_x = M_x^+ \!+\! M_x^-$ measurements
have the same probability. (Here we consider equivalent the measurements that give
the same number of spin ups and downs, independently from permutations---this explains
the combinatorial factor in Eq.~(\ref{PMMS}) and (\ref{PNNMMC}).)

Because the integrals in Eq.~(\ref{PNNMMC}) have the same form as the integral
of Eq.~(\ref{PMMC}), we now have all we need to write down explicitly the result
for Eq.~(\ref{PNNMMC}):
\begin{equation}
P(N_x^+, N_x^- \vert M_x^+, M_x^-) = {{(N_x^+ \! +\! N_x^-)!} \over {N_x^+ ! N_x^- !}} 
{{(M_x^+ \! +\! M_x^- \!+\! 1)! (M_x^+ \! +\! N_x^+)! (M_x^- \! +\! N_x^-)!}
\over{(M_x^+ \! +\! M_x^- \!+\! N_x^+ \! +\! N_x^- \!+\! 1)! M_x^+ ! M_x^- !}}.
\label{PNNMME}
\end{equation}
Eq.~(\ref{PNNMME}) is the exact result, valid for any $N_x^+, N_x^-, M_x^+$ and $M_x^-$.
By the use of the approximation 
\begin{equation}
\lim_{{M}\rightarrow \infty} {{(M \! + \! N)!}\over{M!}} = M^N,
\end{equation}
we can obtain the limit of Eq.~(\ref{PNNMME}) for $M_x \gg N_x$:
\begin{equation}
\lim_{{M_x}\rightarrow \infty} P(N_x^+, N_x^- \vert M_x^+, M_x^-) = 
{{(N_x^+ \! +\! N_x^-)!} \over {N_x^+ ! N_x^- !}} \left({{M_x^+}\over{M_x^+ \! +\! M_x^-}}
\right)^{N_x^+} \left({{M_x^-}\over{M_x^+ \! +\! M_x^-}}\right)^{N_x^-}.
\label{limPNNMM} 
\end{equation}
This is exactly  the result that would follow from using Eq.~(\ref{del?}) in Eq.~(\ref{PNNMMC}).
We have been able to derive it only in the case in which $M_x^+ \! +\! M_x^- \gg N_x^+ , N_x^-$.

As a particular example, the probability of getting $\sigma_x = +1$
$N_x^{+}$ times,
after having observed $\sigma_x = +1$ $M_x^{+}$ times, is:
\begin{equation}
P(N_x^+ \vert M_x^+) = {{\int dx ({1+x\over2})^{M_x^+ \! + \! N_x^+} }
\over {\int dx ({1+x\over2})^{M_x^+} } } = {{M_x^+ + 1}\over{M_x^+ + N_x^+ + 1}}. 
\label{spin_small} 
\end{equation}
(Because for large ${M_x}^+$ the integrals in Eq.~(\ref{spin_small}) are dominated
by a small neighborood of $x=1$, if ${M_x}^+$ is large, the result of Eq.~(\ref{spin_small})
is actually valid for any prior $\Pi(\vec{x})$ that is not zero at $x=1$.) Now, if $M_x^+ \gg N_x^+$,
the probability given by Eq.~(\ref{spin_small}) is 1, as expected for a state that has been
``determined'' to be $\sigma_x = +1$. However, if $M_x^+ = N_x^+$, then---no matter how
large $M_x^+$ is---the probability given by Eq.~(\ref{spin_small}) is only $1\over2$.
This shows the difference of state ``determination'' using Bayes theorem and state
``projection'' from a ``von Neumann'' measurement. A single projective measurement
to the state $\sigma_x = +1$ will force all successive measurements of $\sigma_x$ to
give the result $+1$, no matter how many the successive measurements are.

This ``weakness'' of Bayesian state ``determination'' is of importance for quantum information
theory because such methods are invoked (for instance in Ref.~\cite{Enk02}), to argue that a
conventional laser can be used for quantum teleportation with continuous variables, contrary
to recently made claims (see Ref.~\cite{Rud01}) that teleportation requires novel,
truly coherent light sources.

\section{Laser Beams}
In the rest of this paper, Bayes theorem will be applied to a model of propagating laser beams.
We will discuss the case in which the probability of detecting several photons at a
time is not negligible.
The model for the density operator of two laser beams given in Ref.~\cite{Enk02} is
\begin{equation}
\hat{\Pi}_{\rm{Lasers}} = 
\int_0^{2\pi} {d{\phi_a}\over{2\pi}} \int_0^{2\pi} {d{\phi_b}\over{2\pi}} 
\left(\ket{a e^{i \phi_a}} \bra{a e^{i \phi_a}} \otimes \ket{b e^{i \phi_b}} 
\bra{{b} e^{i \phi_b}}\right)^{\otimes{N}}. \label{laser_beams} 
\end{equation}
The prior of Eq.~(\ref{laser_beams}) follows from a simple symmetry argument:
since there is no reason to prefer any particular phase, choosing an uniform a-priory
distribution is very reasonable. The output beam from each laser is made up by 
$N$ ``packages'', all with the same, unknown phase. (The form of Eq.~(\ref{laser_beams})
also follows from the quantum de Finetti theorem---see Refs.~\cite{Enk02} and \cite{Cav02}.)

A coherent state $\ket{a e^{i\phi}}$ is defined by the eigenvalue condition
\begin{equation}
\hat{a} \ket{a e^{i\phi_a}} = a e^{i\phi_a} \ket{ a e^{i\phi_a}}, \label{cohe1}
\end{equation}
where $\hat{a}$ is a photon destruction operator.
I put hats on all operators: all other symbols are real numbers (except $i$, of course).

\begin{figure}[htbp]
   \centering
   \includegraphics[width=2in]{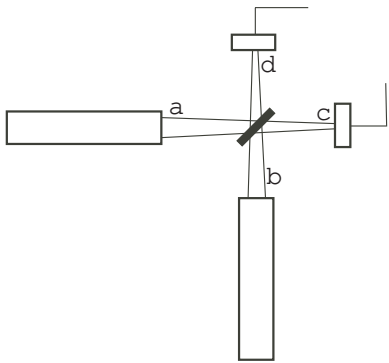} 
   \caption{\label{figure}The output beams from the two lasers (a and b) are mixed on a
   beam splitter
   and the resulting beams (c and d) are measured by two photodetectors.}
\end{figure}

We will measure photons in states obtained by combining the two beams in states produced by a
50/50 lossless splitter/recombiner. The destruction operators of the states produced by the beam
splitter are
\begin{equation}
\hat{c} = {{\hat{a} + \hat{b}}\over\sqrt{2}}
\end{equation}
and
\begin{equation}
\hat{d} = {{\hat{a} - \hat{b}}\over\sqrt{2}}.
\end{equation}
The setup, which is identical to that of Ref.~\cite{Mo:97}, is illustrated in Fig.~\ref{figure}.
The labels c and d will be used to denote both the two beams coming out of the splitter
and the corresponding detectors.
The POVM corresponding to measuring $N_c$ photons at c is
\begin{equation}
\hat{E}_{N_c} = :{{(\eta\hat{c}^{\dagger} \hat{c})^{N_c}}\over{N_c!}}
e^{-\eta\hat{c}^{\dagger}\hat{c}}:, \label{PNFc}
\end{equation}
where the colons represent normal ordering---annihilation operators to the right,
creation operators to the left. (Time ordering is not necessary, because we are considering
a non-interacting field---except for the measurement process.) A similar expression
gives the POVM corresponding to the detection of $N_d$ photons at d:
\begin{equation}
\hat{E}_{N_d} = :{{(\eta\hat{d}^{\dagger} \hat{d})^{N_d}}\over{N_d!}}
e^{-\eta\hat{d}^{\dagger}\hat{d}}:. \label{PNFd}
\end{equation}
Here $\eta$ is a constant, dependent on the quantum efficiency of the detectors.
The time duration of the detection process (which we assume to be short) is included
in the constant $\eta$. (I will not give a proof of Eq.~(\ref{PNFc}) and (\ref{PNFd})---the
reader will find a proof, for instance, in Chapter 14 of Ref.~\cite{Man95}.)
Note that the measurement operators of Eq.~(\ref{PNFc}) and (\ref{PNFd})
consistently take into account the finite efficiency of the measuring devices.
To make the algebra simpler, we assume the same detection efficiency at the two locations.

To provide a normalization, if one measured directly the photons from a---by removing the
beam splitter of Fig.~\ref{figure}, for instance---one would obtain for the probability of
detecting $N_a$ photons
\begin{equation}
P(N_a) = {(\eta a^2)^{N_a} \over {N_a!}} e^{-{\eta a^2}},
\end{equation}
with average number $\eta a^2$. The average number of photons detected
directly from b would be $\eta b^2$.

\subsection{Equal Frequencies}
I will use the Heisenberg picture in these sections. This means the the operators $\hat{E}_{N_c}$,
etc. depend---in general---on time.

I will begin the discussion by assuming that the frequencies of the two lasers are identical.
In this case the time dependence of the operators $\hat{E}_{N_c}$ and $\hat{E}_{N_d}$
can be neglected.  
Inserting Eqs.~(\ref{laser_beams}), (\ref{PNFc}) and (\ref{PNFd}) in Eq.~(\ref{QBPOVMN}),
we have the probability of detecting $N_c$ photons at c in the second package, if $M_c$
photons have been detected at c and $M_d$ photons detected at d in the first package: 
\begin{equation}
P(N_c\vert M_c, M_d) ={{ {\rm{Tr}}( \hat{\Pi}_{\rm{Lasers}}
(\hat{E}_{M_c} \hat{E}_{M_d}) \otimes \hat{E}_{N_c})} \over
{ {\rm{Tr}}( \hat{\Pi}_{\rm{Lasers}} \hat{E}_{M_c} \hat{E}_{M_d})}}, \nonumber
\end{equation}
which gives the result
\begin{equation}
P(N_c\vert M_c, M_d) =  { {\eta^{N_c}}\over{N_c!}}  {{\int_{0}^{2 \pi} {d{\phi}}
({{a^2 \! + \! b^2}\over{2}} \! + \! ab\cos\phi)^{M_c+N_c}
({{a^2 \! + \! b^2}\over{2}} \! - \! ab\cos\phi)^{M_d} e^{-\eta  ({{a^2+b^2}\over{2}}+ab\cos\phi)  }}
 \over {\int_{0}^{2 \pi} {d{\phi}} ({{a^2 \! + \! b^2}\over{2}} \! + \! ab\cos\phi)^{M_c}
({{a^2 \! + \! b^2}\over{2}} \! - \! ab\cos\phi)^{M_d}}}. 
 \label{PNMMF}
\end{equation}
In Eq.~(\ref{PNMMF}) and in the following, $\phi$ denotes the difference of the two
phases: $\phi=\phi_a-\phi_b$. The integral over the sum of the phases does not
affect any of the results, and it will be simply omitted. (The results of Eq.~(\ref{PNMMF})
and successive equations follow from the fact that---because of the normal ordering
and the form  (\ref{laser_beams}) of the density operator---one can substitute the
value ${a e^{i \phi_a} + b e^{i \phi_b}}\over{\sqrt{2}}$ for the $\hat{c}$ operator
and the value ${a e^{i \phi_a} - b e^{i \phi_b}}\over{\sqrt{2}}$ for the $\hat{d}$
operator inside the integral.) The conditions of Eq.~(\ref{QBPOVMN}) are satisfied,
because the measurements on different packages commute, and the measurements
at c and d commute, because $[\hat{d}^{\dagger}, \hat{c}] = [\hat{c}^{\dagger}, \hat{d}] = 0$.
Eq.~(\ref{PNMMF}) can be rewritten as
\begin{equation}
P(N_c\vert M_c, M_d) =  { {\eta^{N_c}}\over{N_c!}}  {{\int_{0}^{2 \pi} {d{\phi}}
P(\phi \vert M_c, M_d)
({{a^2 \! + \! b^2}\over{2}} \! + \! ab\cos\phi)^{N_c} e^{-\eta  ({{a^2+b^2}\over{2}}+ab\cos\phi) }}},
\end{equation}
where $P(\phi \vert M_c, M_d)$ is the conditional probability of the phase (difference) having
the value $\phi$. It is given by the expression
\begin{equation}
P(\phi \vert M_c, M_d) = 
{{({{a^2 \! + \! b^2}\over{2}} \! + \! ab\cos\phi)^{M_c}
({{a^2 \! + \! b^2}\over{2}} \! - \! ab\cos\phi)^{M_d}  }
 \over {\int_{0}^{2 \pi} {d{\phi^{\prime}}} ({{a^2 \! + \! b^2}\over{2}} \! + \! ab\cos\phi^{\prime})^{M_c}
({{a^2 \! + \! b^2}\over{2}} \! - \! ab\cos\phi^{\prime})^{M_d}}}.
\label{PPHIMM}
\end{equation}

In a similar way, we can derive the probability of detecting $N_c$ photons at c and $N_d$
photons at d in the second package, if $M_c$ photons have been detected at c and $M_d$
photons detected at d in the first package: 
\begin{equation}
P(N_c, N_d \vert M_c, M_d) =
{ {\eta^{N_c}}\over{N_c!}}  { {\eta^{N_d}}\over{N_d!}}  {{\int_{0}^{2 \pi} {d{\phi}} ({{a^2 \! + \! b^2}
\over{2}} \! + \! ab\cos\phi)^{M_c+N_c} ({{a^2 \! + \! b^2}\over{2}} \! - \! ab\cos\phi)^{M_d+N_d} e^{-\eta 
({a^2+b^2})  }}\over {\int_{0}^{2 \pi} {d{\phi}} ({{a^2 \! + \! b^2}\over{2}} \! + \! ab\cos\phi)^{M_c}
({{a^2 \! + \! b^2}\over{2}} \! - \! ab\cos\phi)^{M_d}}}. 
 \label{PNNMMF}
\end{equation}
It is reassuring that one can obtain Eq.~(\ref{PNMMF}) by summing Eq.~(\ref{PNNMMF})
over $N_d$.

\subsubsection{Low Intensity}
Eq.~(\ref{PNNMMF}) can be specialized to very low photon detection probabilities---that is,
to the case $\eta {{a^2+b^2}\over{2}} \ll 1$. In this case, we can drop the exponential
and the probabilities of observing one photon at c or d, given the observation
of one photon at c are:
\begin{equation}
P(1, 0 \vert 1, 0) = 
\eta \left({{a^2 + b^2}\over 2} + {{a^2b^2}\over{a^2+b^2}}\right), \label{Pcc}
\end{equation}
\begin{equation}
P(0, 1 \vert 1,  0) = 
\eta \left({{a^2 + b^2}\over 2} - {{a^2b^2}\over{a^2+b^2}}\right). \label{Pdc}
\end{equation}
In particular, if $a=b$, then Eq.~(\ref{Pcc}) and (\ref{Pdc}) imply that,
after the detection of a photon at c, the probability of subsequently detecting
another c photon is 3 times the probability of detecting a d photon.
This, of course, requires that the times between detections be short
compared to the period corresponding to the frequency difference of the two
lasers.

\subsubsection{Multiple Detections}
Multiple detections can determine the phase completely. Using a naive expression
like Eq.~(\ref{del?}), one could argue that after $M_c$ detections at c and $M_d$
detections at d, in the limit of large $M=M_c+M_d$, the probabilities of detecting $N_c$
photons at c or $N_d$ photons at d are:
\begin{equation}
\lim_{{M_c+M_d} \rightarrow \infty} P(N_c\vert M_c, M_d) = 
{ {\eta^{N_c}}\over{N_c!}}  \left({{a^2+b^2}\over{2}}+ab\cos\phi_0\right)^{N_c} 
e^{-\eta ({{a^2+b^2}\over{2}}+ab\cos\phi_0)}  \label{PcNN} 
\end{equation}
and
\begin{equation}
\lim_{{M_c+M_d} \rightarrow \infty} P(N_d\vert M_c, M_d) =   
{ {\eta^{N_d}}\over{N_d!}}  \left({{a^2+b^2}\over{2}}-ab\cos\phi_0\right)^{N_d} 
e^{-\eta  ({{a^2+b^2}\over{2}}-ab\cos\phi_0)}, \label{PdNN} 
\end{equation}
where $\cos\phi_0$ is given by the expression
\begin{equation}
\cos\phi_0 = \left({{a^2+b^2}\over{2ab}}{{M_c-M_d}\over{M_c+M_d}}\right). \label{Phi0}
\end{equation}
Note that Eq.~(\ref{Phi0}) determines $\phi_0$ only up to a sign, but---since Eqs.~(\ref{PcNN})
and (\ref{PdNN}) only depend on  $\cos\phi_0$--- the results are unambiguous.
The result of Eq.~(\ref{Phi0}) is only obtained if the detection efficiency is the same at c and d.
The probabilities of Eqs.~(\ref{PcNN}) and (\ref{PdNN}) are the same results that would
follow from coherent states with a fixed phase difference $\phi_0$.
Effectively, the multiple detections have determined the phase (difference) completely.
This is an illustration of what Ref.~\cite{Enk02} calls a ``phase-lock without phase''.

One could worry that Eq.~(\ref{Phi0}) gives a complex phase
if $\left|{{M_c-M_d}\over{M_c+M_d}}\right| > {{2ab}\over{a^2+b^2}}$.
However, the a-priori probability of detecting $M_c$ and $M_d$ photons is
\begin{equation}
P(M_c, M_d) ={{\eta^{M_c}}\over{M_c!}} { {\eta^{M_d}}\over{M_d!}}
{{\int_0^{2\pi} {{d{\phi}}\over{2\pi}}({{a^2+b^2}\over{2}}+ab\cos\phi)^{M_c}
({{a^2+b^2}\over{2}}-ab\cos\phi)^{M_d} e^{-\eta  ({a^2+b^2})  }}}.    \label{PMMF}
\end{equation}
Now, the function to be integrated in Eq.~(\ref{PMMF}) has, for large $M_c$ and $M_d$,
sharp peaks at $\cos\phi_0$ given in Eq.~(\ref{Phi0}) and is neglegible elsewhere.
So the probability of Eq.~(\ref{PMMF}) will vanish for large $M_c$, $M_d$,
unless $\left|{{M_c-M_d}\over{M_c+M_d}}\right| \le {{2ab}\over{a^2+b^2}}$,
because otherwise the peaks would be outside the integration region.

It should be obvious that it is possible to rigorously derive the results of
Eqs.~(\ref{PcNN}) and (\ref{PdNN}) only if ${M_c} + {M_d}\gg{N_c}, {N_d}$.
Otherwise, Eq.~(\ref{PNMMF}) represent the only prediction for the future detections
that follows from the initial observation. This means that it is not entirely correct to state
as in Ref.~\cite{Enk02} that ``appropriate measurements of part of a laser beam
will reduce the quantum state of the rest of the laser beam to a pure coherent state''.
The phase measurement on the first package will ``reduce'' only the first package:
the phase of the rest of the beams will be ``determined'' only in a statistical way. 
Again, this is a ``weakness'' of the Bayesian methods used. In practice, however,
a statistical state reconstruction could be sufficiently accurate for practical
applications, like quantum cryptography, so the conclusions of Ref.~\cite{Enk02}
are still correct in practice.

\subsection{Different Frequencies}
I will now briefly discuss the case in which the two lasers have different frequencies.
Up to this point, we have assumed that the time length of the photon detections and
the time separation of initial and final detections are short with respect to the period
corresponding to the difference of the frequencies of the two lasers.
If only the length of the detection processes is short, but the time difference
between the photon detections is not, then the probability of detecting
$N_c$ photons at c, at time $t$---after detecting $M_c$ photons at c and $M_d$
photons at d, at time $0$---is
\begin{equation}
P(N_c\vert M_c, M_d) =    { {\eta^{N_c}}\over{N_c!}}
{{\int_{0}^{2 \pi} {d{\phi}} P(\phi \vert M_c, M_d) 
({{a^2 \! +\! b^2}\over{2}} \! + \!
ab\cos(\phi \! -\! \Delta{\omega} t))^{N_c} e^{-\eta  ({{a^2+b^2}\over{2}}+
ab\cos(\phi-\Delta{\omega} t))  }}},   \label{PNMMT} 
\end{equation}
where $\Delta{\omega}$ is the frequency difference:
\begin{equation}
\Delta{\omega} = \omega_a - \omega_b,  \label{domega}
\end{equation}
and $P(\phi \vert M_c, M_d)$ is given by Eq.~(\ref{PPHIMM})
(Eq.~(\ref{PNMMT}) is derived by substituting the value
${a e^{i \phi_a - i \omega_a t} + b e^{i \phi_b - i \omega_b t}}\over{\sqrt{2}}$ for $\hat{c}(t)$,
etc.---substitutions made possible by the normal ordering.)  

Because of the ambiguity in the sign of $\phi$ in Eq.~(\ref{PPHIMM}), measurements at different
times are necessary to get a complete picture.
If one earlier detects ${M_c}(t_i)$ photons at c, and ${M_d}(t_i)$ photons at d at time $t_i$---where
$i=0,1,...m$---the
conditional probability of detecting ${N_c}$ photons at c, at time $t$ is
\begin{eqnarray}
\lefteqn{P(N_c\vert {M_c}(t_i), {M_d}(t_i)) = } \nonumber \\
& & { {\eta^{N_c}}\over{N_c!}}
\int_{0}^{2 \pi} \! {d{\phi}} P(\phi \vert {M_c}(t_i), {M_d}(t_i)) ({{a^2 \! +\! b^2}\over{2}} \! + \!
ab\cos(\phi \! -\! \Delta{\omega} t))^{N_c} e^{-\eta  ({{a^2+b^2}\over{2}}+
ab\cos(\phi-\Delta{\omega} t))  },  \label{PNMMTN} 
\end{eqnarray}

were $P(\phi \vert {M_c}(t_i), {M_d}(t_i))$, the conditional probability of the phase having the
value $\phi$, after the measurements at times $t_0, t_1 \dots t_m$, is given by the expression
\begin{eqnarray}
\lefteqn{ P(\phi \vert {M_c}(t_i), {M_d}(t_i)) = } \nonumber \\
& & {{\prod_{i=0}^{m} \left( ({{a^2 \! + \! b^2}\over{2}} \! + \! ab\cos(\phi \! - \! \Delta{\omega} t_i))^{M_c(t_i)}
({{a^2 \! + \! b^2}\over{2}} \! - \! ab\cos(\phi \! - \! \Delta{\omega} t_i))^{M_d(t_i)} \right)}
\over{\int_{0}^{2 \pi} {d{\phi^{\prime}}} \prod_{i=0}^{m} \left( ({{a^2 \! +\! b^2}\over{2}} \! + \! ab\cos(\phi^{\prime}
\! - \! \Delta{\omega} t_i))^{M_c(t_i)}
({{a^2 \! + \! b^2}\over{2}} \! - \! ab\cos(\phi^{\prime} \! -\! \Delta{\omega} t_i))^{M_d(t_i)} \right)}}.
\label{PPHIMMN}
\end{eqnarray}

Eqs.~(\ref{PNMMTN}) and (\ref{PPHIMMN})
describe how coherent oscillations can be predicted by earlier measurements. 
They give an analytic description of the process of ``generating'' coherent oscillations
from the entanglement produced by photon detections, as described by Klaus
M{\o}lmer in Ref.~\cite{Mo:97}.  Eqs.~(\ref{PNMMTN}) and (\ref{PPHIMMN}) also provide
a quantitative prediction from the model of Ref.~\cite{Enk02} for propagating laser beams,
a prediction that could be used to test the model.

\section{Conclusion}
The discussion presented in the previous section is, hopefully, a contribution
to the continuing discussion of the problem of quantum phase.
Differently from other authors (see, for instance, Ref.~\cite{Hradil95}),
I did not try to define a phase ``observable''.
In my approach, phase is a parameter of an (initially) unknown
quantum state. Photon detections determine a probability distribution
for the unknown phase.

The information-theoretic approach to quantum phase used in this paper has 
a precedent in Ref.~\cite{Hradil96} (which, however, dealt with neutron interferometry).
The approach used in the present paper does not make use of semiclassical
methods and takes into account both photon statistics
and the finite efficiency of the detectors.

\bibliography{NeriQB7}

\begin{thebibliography}{10}
\expandafter\ifx\csname natexlab\endcsname\relax\def\natexlab#1{#1}\fi
\expandafter\ifx\csname bibnamefont\endcsname\relax
  \def\bibnamefont#1{#1}\fi
\expandafter\ifx\csname bibfnamefont\endcsname\relax
  \def\bibfnamefont#1{#1}\fi
\expandafter\ifx\csname citenamefont\endcsname\relax
  \def\citenamefont#1{#1}\fi
\expandafter\ifx\csname url\endcsname\relax
  \def\url#1{\texttt{#1}}\fi
\expandafter\ifx\csname urlprefix\endcsname\relax\def\urlprefix{URL }\fi
\providecommand{\bibinfo}[2]{#2}
\providecommand{\eprint}[2][]{\url{#2}}

\bibitem[{\citenamefont{Schack et~al.}(2001)\citenamefont{Schack, Brun, and
  Caves}}]{Sch:01}
\bibinfo{author}{\bibfnamefont{R.}~\bibnamefont{Schack}},
  \bibinfo{author}{\bibfnamefont{T.~A.} \bibnamefont{Brun}}, \bibnamefont{and}
  \bibinfo{author}{\bibfnamefont{C.~M.} \bibnamefont{Caves}},
  \bibinfo{journal}{Phys.\ Rev.\ A} \textbf{\bibinfo{volume}{64}},
  \bibinfo{pages}{014305} (\bibinfo{year}{2001}).

\bibitem[{\citenamefont{Bu\u{z}ek et~al.}(1998)\citenamefont{Bu\u{z}ek, Derka,
  Adam, and Knight}}]{Buz:98}
\bibinfo{author}{\bibfnamefont{V.}~\bibnamefont{Bu\u{z}ek}},
  \bibinfo{author}{\bibfnamefont{R.}~\bibnamefont{Derka}},
  \bibinfo{author}{\bibfnamefont{G.}~\bibnamefont{Adam}}, \bibnamefont{and}
  \bibinfo{author}{\bibfnamefont{P.~L.} \bibnamefont{Knight}},
  \bibinfo{journal}{Ann.\ Phys.} \textbf{\bibinfo{volume}{266}},
  \bibinfo{pages}{454} (\bibinfo{year}{1998}).

\bibitem[{\citenamefont{Hudson and Moody}(1976)}]{Hud76}
\bibinfo{author}{\bibfnamefont{R.~L.} \bibnamefont{Hudson}} \bibnamefont{and}
  \bibinfo{author}{\bibfnamefont{G.~R.} \bibnamefont{Moody}},
  \bibinfo{journal}{Z.\ Wahrs.} \textbf{\bibinfo{volume}{33}},
  \bibinfo{pages}{343} (\bibinfo{year}{1976}).

\bibitem[{\citenamefont{Caves et~al.}(2002)\citenamefont{Caves, Fuchs, and
  Schack}}]{Cav02}
\bibinfo{author}{\bibfnamefont{C.~M.} \bibnamefont{Caves}},
  \bibinfo{author}{\bibfnamefont{C.}~\bibnamefont{Fuchs}}, \bibnamefont{and}
  \bibinfo{author}{\bibfnamefont{R.}~\bibnamefont{Schack}},
  \bibinfo{journal}{J.\ Math.\ Phys} \textbf{\bibinfo{volume}{43}},
  \bibinfo{pages}{4537} (\bibinfo{year}{2002}).

\bibitem[{\citenamefont{van Enk and Fuchs}(2002)}]{Enk02}
\bibinfo{author}{\bibfnamefont{S.~J.} \bibnamefont{van Enk}} \bibnamefont{and}
  \bibinfo{author}{\bibfnamefont{C.~A.} \bibnamefont{Fuchs}},
  \bibinfo{journal}{Phys.\ Rev.\ Lett.} \textbf{\bibinfo{volume}{88}},
  \bibinfo{pages}{027902} (\bibinfo{year}{2002}).

\bibitem[{\citenamefont{Rudolph and Sanders}(2001)}]{Rud01}
\bibinfo{author}{\bibfnamefont{T.}~\bibnamefont{Rudolph}} \bibnamefont{and}
  \bibinfo{author}{\bibfnamefont{B.~C.} \bibnamefont{Sanders}},
  \bibinfo{journal}{Phys.\ Rev.\ Lett.} \textbf{\bibinfo{volume}{87}},
  \bibinfo{pages}{077903} (\bibinfo{year}{2001}).

\bibitem[{\citenamefont{M{\o}lmer}(1997)}]{Mo:97}
\bibinfo{author}{\bibfnamefont{K.}~\bibnamefont{M{\o}lmer}},
  \bibinfo{journal}{Phys.\ Rev.\ A} \textbf{\bibinfo{volume}{55}},
  \bibinfo{pages}{3195} (\bibinfo{year}{1997}).

\bibitem[{\citenamefont{Mandel and Wolf}(1995)}]{Man95}
\bibinfo{author}{\bibfnamefont{L.}~\bibnamefont{Mandel}} \bibnamefont{and}
  \bibinfo{author}{\bibfnamefont{E.}~\bibnamefont{Wolf}},
  \emph{\bibinfo{title}{Optical Coherence and Quantum Optics}}
  (\bibinfo{publisher}{Cambridge Univ. Press}, \bibinfo{address}{Cambridge},
  \bibinfo{year}{1995}).

\bibitem[{\citenamefont{{Hradil}}(1995)}]{Hradil95}
\bibinfo{author}{\bibfnamefont{Z.}~\bibnamefont{{Hradil}}},
  \bibinfo{journal}{\pra} \textbf{\bibinfo{volume}{51}}, \bibinfo{pages}{1870}
  (\bibinfo{year}{1995}).

\bibitem[{\citenamefont{{Hradil} et~al.}(1996)\citenamefont{{Hradil}, {My{\v
  s}ka}, {Pe{\v r}ina}, {Zawisky}, {Hasegawa}, and {Rauch}}}]{Hradil96}
\bibinfo{author}{\bibfnamefont{Z.}~\bibnamefont{{Hradil}}},
  \bibinfo{author}{\bibfnamefont{R.}~\bibnamefont{{My{\v s}ka}}},
  \bibinfo{author}{\bibfnamefont{J.}~\bibnamefont{{Pe{\v r}ina}}},
  \bibinfo{author}{\bibfnamefont{M.}~\bibnamefont{{Zawisky}}},
  \bibinfo{author}{\bibfnamefont{Y.}~\bibnamefont{{Hasegawa}}},
  \bibnamefont{and} \bibinfo{author}{\bibfnamefont{H.}~\bibnamefont{{Rauch}}},
  \bibinfo{journal}{Physical Review Letters} \textbf{\bibinfo{volume}{76}},
  \bibinfo{pages}{4295} (\bibinfo{year}{1996}).

\end{thebibliography}

\end{document}